\documentclass[
 amsmath,amssymb,
 aps, physrev,
reprint
]{revtex4-2}

\usepackage{graphicx}
\usepackage{dcolumn}
\usepackage{bm}
\usepackage{color}
\usepackage[colorlinks=true,
            citecolor=blue,
            linkcolor=blue,
            urlcolor=blue]{hyperref}

\begin{document}

\title{\textbf{Autoencoder-Based Unsupervised Identification of Nonequilibrium Phases in Sheared Binary Colloids} 
}%

\author{Yoshitaka Miyahara}
\email[]{sb22074r@st.omu.ac.jp}

\author{Taiki Haga}
\email[]{taiki.haga@omu.ac.jp}

\affiliation{Department of Physics and Electronics, Osaka Metropolitan University, Sakai-shi, Osaka 599-8531, Japan}

\date{\today}

\begin{abstract}
Identifying nonequilibrium phases in particle systems remains a major challenge because they often exhibit complex and spatially heterogeneous structures without long-range order.
Here, we develop an unsupervised machine-learning framework for classifying such nonequilibrium phases by integrating Fourier-based preprocessing, an autoencoder, and a Gaussian mixture model (GMM).
Specifically, we transform global spatial configurations into Fourier space and use the amplitudes of Fourier coefficients as inputs to the autoencoder.
This preprocessing suppresses spatial noise while preserving phase-specific structural features and physical interpretability.
We demonstrate the effectiveness of this framework using a binary charged colloidal system under steady shear flow, where the competition between Coulomb interactions and shear gives rise to three nonequilibrium phases characterized by distinct local structures.
The encoded latent space reveals well-separated clusters that are robustly identified by the GMM, enabling the construction of a nonequilibrium phase diagram based on cluster membership probabilities.
The resulting phase boundaries are consistent with those independently obtained from radial distribution function analysis and unsupervised anomaly detection.
These results demonstrate that autoencoder-based unsupervised learning provides an effective framework for identifying nonequilibrium phases in complex particle systems.
\end{abstract}

\maketitle

{\em Introduction.--}
Many recent studies have demonstrated that unsupervised machine learning is highly effective in identifying equilibrium phases of matter \cite{Wang2016, Wetzel2017, Hu2017, Mendes2021, Rodriguez2019, Scheurer2020, Yu2020, Alexandrou2020, Frk2025, Jadrich2018}. Such methods allow us to classify the phases of complex systems without prior theoretical knowledge. The fundamental approach involves compressing high-dimensional data, generated from microscopic configurations of the system, into a low-dimensional latent space using techniques such as principal component analysis or autoencoders. While previous studies have primarily validated these techniques using simple models with well-understood phase structures, there is a growing demand to extend these applications to explore previously unknown or highly complex systems.

Identifying phase structures in nonequilibrium steady states presents a significantly greater challenge compared to equilibrium systems due to their vast diversity and complexity \cite{Hinrichsen2000, Acharyya2005, Cate2015, Liu2010, Vicsek1995}. For instance, colloidal suspensions are known to form a variety of intricate patterns when subjected to shear flow \cite{Vermant2005, Besseling2012, Hecht2006, Ghimenti2024, Ruiz2018}. Generally, driving forces such as shear tend to destabilize the globally ordered structures that emerge in equilibrium. In certain two-dimensional (2D) monodisperse colloidal systems, shear has been shown to induce a two-stage melting process, where a crystalline phase transforms into a liquid phase via an intermediate hexatic phase \cite{Delhommelle2004}. Furthermore, 2D colloidal clusters that form hexagonal lattices in equilibrium exhibit string-like arrangements when subjected to shear \cite{Stancik2003}. Despite these extensive studies, understanding what types of disordered or locally ordered nonequilibrium phases arise from distinct equilibrium crystal structures under shear remains an open problem.

In this Letter, we propose an unsupervised machine learning framework to classify disordered nonequilibrium phases of particle systems. Because the direct application of autoencoders to raw configurations often fails to capture essential characteristics, we preprocess the spatial data via Fourier transformation and use the resulting amplitudes as inputs for dimensionality reduction. Subsequently, we apply a Gaussian mixture model (GMM) to the latent space to identify clusters corresponding to distinct physical phases \cite{Boattini2019, Boattini2020}. 

\begin{figure*}
\centering
\includegraphics[width=\textwidth]{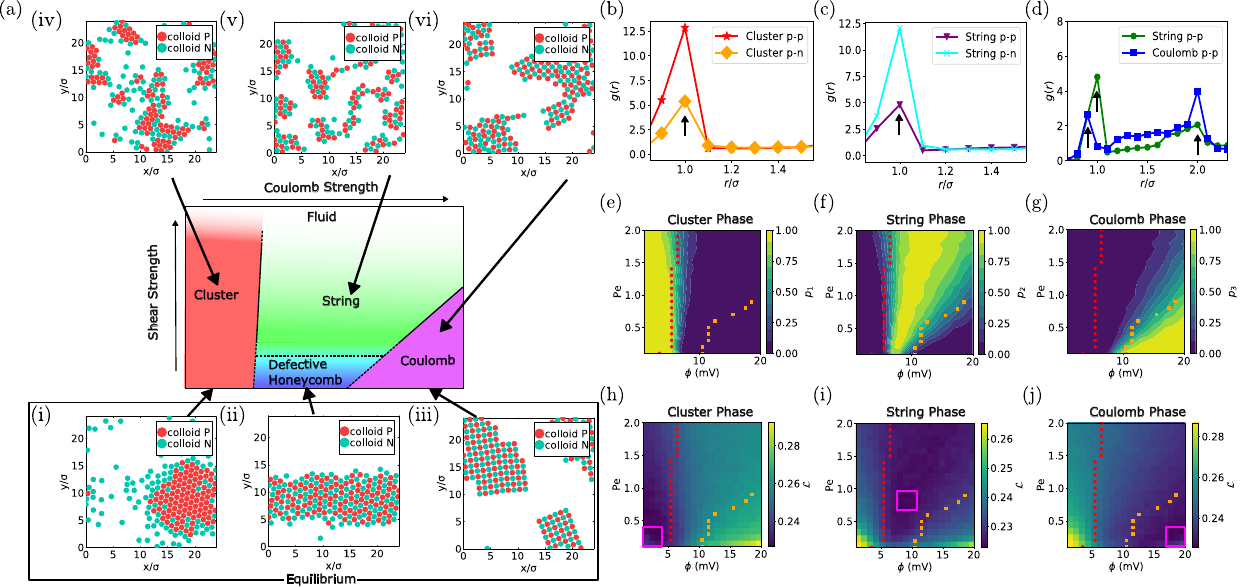}
\caption{(a) Schematic phase diagram as a function of the shear strength $Pe$ and Coulomb interaction $\phi$. The insets (i)-(vi) show representative snapshots of colloidal cconfigurations, where (i)-(iii) and (iv)-(vi) correspond to $Pe=0$ (equilibrium) and $Pe=0.1$, respectively.
Within each row, the panels are arranged from left to right with increasing Coulomb interaction strength, $\phi = 3$, $7$, and $18~\mathrm{mV}$.
(b), (c) Comparison of the RDFs for the Cluster and String phases at $Pe=0.5$, corresponding to $\phi=3 \: \text{mV}$ and $8 \: \text{mV}$, respectively.
In the Cluster phase, the first peak of $g_{pp}(r)$ at $r/\sigma = 1.0$ (indicated by a black arrow) exceeds that of $g_{pn}(r)$, whereas the opposite trend is observed in the String phase.
(d) Comparison of \(g_{pp}(r)\) in the String and Coulomb phases at $Pe=0.5$, corresponding to $\phi=8 \: \text{mV}$ and $18 \: \text{mV}$, respectively.
In the String phase, the first peak of $g_{pp}(r)$ is higher than its second peak marked by a black arrow, and this relationship is reversed in the Coulomb phase. 
(e)-(g) Heat maps showing the probability $p_i$ belonging to each
phase obtained from our method, plotted as a function of $Pe$ and $\phi$. 
The scattered points indicate the phase boundaries determined from the RDF analysis; red circles and orange squares correspond to the Cluster–String and
String–Coulomb boundaries, respectively. 
(h)-(j) Heat maps of the reconstruction loss obtained from the anomaly
detection method, with pink square frames indicating the training regions in parameter space.}
\label{fig:phase_RDF}
\end{figure*}

We demonstrate the effectiveness of our framework on a testbed system of 2D binary charged colloids under shear flow, investigated via Brownian dynamics simulations. At equilibrium, this system exhibits three types of crystalline phases. Although the application of shear flow destroys this global order, our proposed method robustly identifies three nonequilibrium phases characterized by distinct local structures [see Fig.~\ref{fig:phase_RDF}(a)]. Furthermore, we show that the nonequilibrium phase diagram constructed using our machine learning approach is consistent with the phase boundaries estimated from radial distribution functions (RDFs). Finally, we implement an anomaly detection method based on the autoencoder's reconstruction loss, providing an independent validation of our primary clustering results \cite{Kottmann2020, Acevedo2021, Kaming2021, Kottmann2021}.

{\em Binary colloidal system under shear.--}
We consider a 2D system of oppositely charged colloids (denoted as P and N) under linear shear. A colloid P (N) is positively (negatively) charged, and both types of colloids have a diameter $\sigma$ but are composed of different materials. 
The interaction between colloids is described by a Derjaguin-Landau-Verwey-Overbeek (DLVO)-like potential, $U_{\text{DLVO}} = U_{\text{el}} + U_{\text{vdW}}$, which consists of a screened Coulomb potential $U_{\text{el}}$ and a van der Waals potential $U_{\text{vdW}}$ \cite{Verwey1947}.
The screened Coulomb interaction between two colloids at center-to-center distance $r_{ij} = |\bm{r}_i - \bm{r}_j|$ is given by 
$$U_{\text{el}}(r_{ij}) = \pi\epsilon\sigma\phi_i\phi_j\left(1 + \frac{2\lambda}{\sigma}\right)^2\exp\left(-\frac{r_{ij} - \sigma}{\lambda}\right),$$
where $\epsilon$ represents the solvent permittivity, $\phi_i$ is the electrostatic surface potential of the $i$th colloid ($\phi_i = \pm \phi$ with $\phi > 0$), and $\lambda$ is the Debye screening length, which is set to $\lambda = \sigma/4$.

The van der Waals interaction $U_{\text{vdW}}(r_{ij})$ can be expressed as 
$$U_{\text{vdW}}(r_{ij}) = -\frac{A_{ij}\sigma}{24(r_{ij} - \sigma)},$$
with the Hamaker constant $A_{ij}$, which depends on the materials of the colloids \cite{Israelachvili2011}.
In this study, we set the van der Waals interaction between P-type colloids to be stronger than that for P-N or N-N pairs.

We treat colloids as nearly hard spheres using the Weeks-Chandler-Andersen (WCA) potential $U_{\text{WCA}}(r_{ij})$, such that the total interaction is given by $U = U_{\text{WCA}} + U_{\text{DLVO}}$ \cite{Weeks1971}.
Using this potential, we investigate the nonequilibrium patterns of the colloids at room temperature under linear shear flow via Brownian dynamics simulation. 
The dynamics of the system are described by the following overdamped Langevin equation:
$$\xi \left[ \frac{d\bm{r}_i}{dt} + \gamma(y_i - L/2)\bm{\hat{e}}_x \right] = -\sum_{j (\neq i)}\nabla_i U(r_{ij}) + \bm{f}_i^{\text{ran}},$$
where $\xi$ is the friction coefficient, $\bm{r}_i = (x_i, y_i)$ is the position of the $i$th colloid, $\nabla_i = (\partial/\partial x_i, \partial/\partial y_i)$, and $\gamma$ denotes the shear rate. 
The term $\bm{f}_i^{\text{ran}}(t) = (f_{i, x}^{\text{ran}}(t), f_{i, y}^{\text{ran}}(t))$ is the Gaussian random force acting on the $i$th colloid, which satisfies $\langle f_{i,\alpha}^{\text{ran}}(t) \rangle = 0$ and $\langle f_{i,\alpha}^{\text{ran}}(t) f_{j,\beta}^{\text{ran}}(t') \rangle = 2\xi k_B T\delta_{ij} \delta_{\alpha \beta} \delta(t-t')$.
The dimensionless shear rate is characterized by the Peclet number $Pe$ defined as
$$Pe = \frac{\gamma \xi \sigma^2}{6\pi k_B T}.$$

The simulations are conducted in a square domain of size $L = 24 \sigma$, containing 105 colloids of each type, resulting in a total number $N_{\text{tot}} = 210$.
We impose periodic boundary conditions along the x-axis and Lees–Edwards boundary conditions along the y-axis by introducing linear shear via sliding periodic images \cite{Lees1972}.
Further details of the system and simulation protocol are provided in the Supplemental Material \cite{SupMat}. 

{\em Phases.--}
Figure~\ref{fig:phase_RDF}(a) shows the schematic phase diagram of our system, based on the observations, and the representative configurations in each phase.
Our aim is to clarify how shear modifies the phase behavior and to construct a physically consistent nonequilibrium phase diagram in the $(\phi,Pe)$ parameter space using machine learning.
First, we identify the equilibrium phases in the absence of shear ($Pe=0$) by gradually cooling the system from a high-temperature disordered state.
The detailed annealing protocol is described in the Supplemental Material \cite{SupMat}.
In equilibrium, increasing the Coulomb force leads to three distinct phases: Cluster, Defective Honeycomb, and Coulomb phases corresponding to the insets (i), (ii), (iii) in Fig.~\ref{fig:phase_RDF}(a), respectively.
As $\phi$ increases, the morphology of P-colloid aggregates evolves from a gigantic cluster into a triangular lattice with partial honeycomb order.
At sufficiently large $\phi$, oppositely charged colloids alternate to form an ionic-crystal-like structure.

The configurations in the nonequilibrium phases under shear are shown in the insets (iv), (v), (vi) in Fig.~\ref{fig:phase_RDF}(a), where increasing the shear rate triggers the fragmentation of P-colloid aggregates into locally ordered, spatially inhomogeneous clusters.
At low $\phi$, these smaller aggregates remain robust against shear, so this region is still classified as the Cluster phase [inset (iv)]. 
At intermediate $\phi$, string-like structures emerge, replacing the partial honeycomb order observed in equilibrium [inset (v)].
We distinguish this state from the Defective Honeycomb phase and refer to it as the String phase.
At high $\phi$, the Coulomb phase retains its ionic-crystal-like structure, despite the presence of defects composed of string-like arrangements of P and N colloids [inset (vi)].
As $Pe$ increases further, these string-like patterns become more prominent, while the ionic-crystal-like order is progressively disrupted.
This shear-induced disruption drives the crossover from the Coulomb phase to the String phase, as shown in the phase diagram in Fig.~\ref{fig:phase_RDF}(a).

As shown in Fig.~\ref{fig:phase_RDF}(a), increasing the shear rate makes it more difficult to distinguish between different phases by visual inspection.
To establish a baseline for phase identification, we first determine the phase boundaries using a traditional metric: the radial distribution function (RDF).
Here, we focus on two types of RDFs $g_{pp}(r)$ and $g_{pn}(r)$, which represent the density of colloids P and N, respectively, at a distance $r$ from a reference colloid P.
The Cluster phase is characterized by the first peak of $g_{pp}(r)$ being higher than that of $g_{pn}(r)$ [Fig.~\ref{fig:phase_RDF}(b)], whereas this relationship is reversed in the String phase [Fig.~\ref{fig:phase_RDF}(c)].
In the Coulomb phase, the ionic-crystal-like order is reflected in $g_{pp}(r)$ such that the first peak of $g_{pp}(r)$ at contact is lower than the second peak around $r = 2\sigma$ [Fig.~\ref{fig:phase_RDF}(d)].
Based on these criteria, we determine the phase boundaries, shown as the scattered points in Figs.~\ref{fig:phase_RDF}(e)-(j).
In the following, we demonstrate that these boundaries derived from the human-crafted physical characteristics coincide with those obtained via our unsupervised learning method.

\begin{figure}
\centering
\includegraphics[width=\columnwidth]{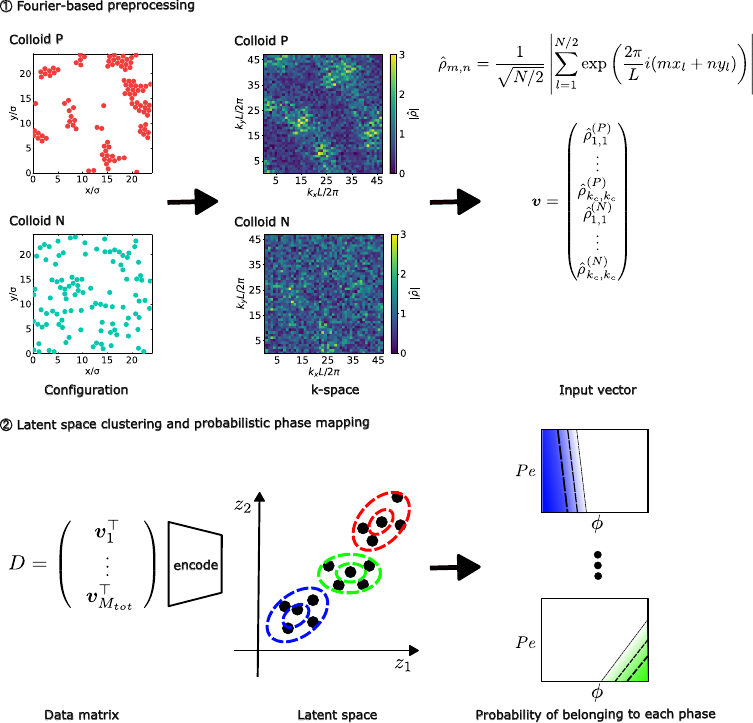}
\caption{Schematic illustration of our method.
First, we calculate the amplitudes of the Fourier coefficients from sampled configurations and assemble them into feature vectors.
After repeating this procedure for all configurations, dimensionality reduction is performed using an autoencoder, and the encoded latent space is clustered with a GMM.
Finally, the phase diagram is constructed from the probabilities that data points at each parameter set belong to each cluster.}
\label{fig:Method}
\end{figure}

{\em Method.--}
Our methodology integrates an autoencoder for dimensionality reduction with a GMM to identify clusters within the latent space, which correspond to the different physical phases.
While previous studies have demonstrated that the autoencoder-based dimensionality reduction is effective to classify various local motifs in colloidal systems when applied to local structural descriptors \cite{Boattini2019, Boattini2020}, phase identification in our system necessitates the inclusion of global statistical information.
However, we find that applying the autoencoder directly to raw configurations is ineffective, as it fails to reconstruct highly disordered configurations in its decoding process (see the Supplemental Material \cite{SupMat}).

To overcome this limitation, we preprocess each colloidal configuration by Fourier transformation and use the amplitudes of the Fourier coefficients as inputs to the autoencoder.
For each colloidal species, the Fourier amplitude is defined as

\begin{equation}
\hat{\rho}_{m,n} = 
\frac{1}{\sqrt{N/2}} \left| \sum_{l=1}^{N/2} \exp \left( i \frac{2\pi}{L} (mx_l+ny_l) \right) \right|,
\end{equation}
where $(x_l,y_l)$ denotes the position of the $l$th colloid and $(m,n)$ specifies the wave vector.
Only Fourier modes satisfying $1\le{}m,n\le n_c$ are retained, with $n_c=48$, corresponding to a minimum resolved length scale of $\sigma/2$.
The resulting Fourier amplitudes are assembled into a feature vector and used as the input to the autoencoder.
Although taking the absolute values of these coefficients involves some loss of information, it eliminates the degrees of freedom associated with global translations of the system.
This approach effectively enhances phase-specific features while preserving physical interpretability without requiring any tunable hyperparameters.

The procedure of our method is summarized below and illustrated in Fig.~\ref{fig:Method}:
1. Generate $M$ configurations with different initial conditions for fixed values of $\phi$ and $Pe$ using Brownian dynamics simulations.
2. Repeat step 1 across various combinations of $\phi$ and $Pe$ to collect a total of $M_{\text{tot}}$ configurations.
3. For each configuration, compute the amplitudes of the Fourier coefficients and assemble them into a single feature vector.
4. Perform dimensionality reduction using the autoencoder.
5. Apply a GMM to identify clusters within the latent space and assign the probability of each configuration belonging to each cluster.

The autoencoder consists of input and output layers with 4608 nodes, corresponding to the dimensionality of the input data $2 n_c^2$, two hidden layers with 128 nodes each, and a two-dimensional latent layer.
For each parameter set $(\phi,Pe)$, 100 configurations are used for training, resulting in a total training dataset of $M_{tot}=40,000$ samples.

\begin{figure*}
\centering
\includegraphics[width=\textwidth]{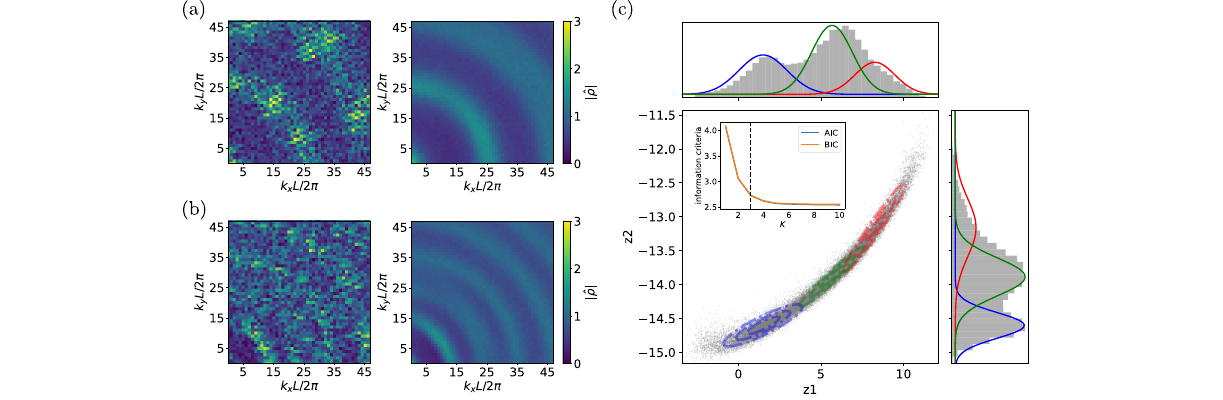}
\caption{(a), (b) Comparison of the original and reconstructed Fourier data at a fixed $Pe = 0.1$ for $\phi = 1 \: \text{mV}$ and $20 \: \text{mV}$, respectively.
(c) Latent space representation shown as scatter points, accompanied by one-dimensional histograms projected onto each axis. 
The colored contours in the main plot and solid curves in the projected histograms denote the Gaussian components derived from the GMM. 
The inset shows the Akaike and Bayesian information criteria (AIC and BIC) as a function of the number of clusters $K$.}
\label{fig:reconstruction_latent}
\end{figure*}

{\em Result and discussion.--}
Figure~\ref{fig:reconstruction_latent} presents the results of the dimensionality reduction of Fourier-transformed data using the autoencoder.
To demonstrate that the autoencoder successfully captures essential features, we compare the original and reconstructed configurations of colloids P in Fourier space at $Pe = 0.1$ for $\phi = 1 \: \text{mV}$ and $20 \: \text{mV}$ [Figs.~\ref{fig:reconstruction_latent}(a) and (b), respectively].
In Fig.~\ref{fig:reconstruction_latent}(a), the original data exhibit broad peaks along the circle $|k| \approx 24 \times (2\pi / L)$, corresponding to the average interparticle distance within dense clusters of colloids P.
The reconstructed data faithfully reproduce these features as smoothed rings.
The same behavior is observed in Fig.~\ref{fig:reconstruction_latent}(b), where the autoencoder captures narrow peaks at $|k| \approx 12 \times (2\pi / L)$.
These peaks reflect the ionic-crystal-like structure, where the characteristic distance between colloids P is larger than that in the Cluster phase.
These results highlight the advantage of Fourier preprocessing over raw configurations.
The Fourier representation effectively averages out spatial inhomogeneities and encodes structural information into distinct peaks at characteristic wavenumbers, which are easily reproduced by the autoencoder.
Additional comparisons across the full parameter space, as well as the training loss over epochs, are provided in the Supplemental Material \cite{SupMat}.

The resulting latent space is shown in Fig.~\ref{fig:reconstruction_latent}(c), where the encoded Fourier data are represented as gray scatter points, together with one-dimensional histograms projected onto each axis.
The data points are distributed along a continuous curve, exhibiting two distinct peaks in their density distribution. 
To identify the underlying structure, we apply a GMM while varying the number of clusters $K$. 
The inset of Fig.~\ref{fig:reconstruction_latent}(c) shows the Akaike and Bayesian information criteria (AIC and BIC), where lower values indicate a more optimal choice of $K$ \cite{Akaike1974, Schwarz1978}.
While neither criterion exhibits a clear minimum, both show a kink at $K = 3$, marking a crossover from a steep to a gradual decrease.
Based on this observation, we visualize the GMM components for $K = 3$ as contours in Fig.~\ref{fig:reconstruction_latent}(c) and as solid curves in the histograms. 
By comparing these components with the phase diagram, we find that the blue, green, and red clusters correspond to the Cluster, String, and Coulomb phases, respectively.
The decomposition of the broader asymmetric peak into two Gaussian components (green and red) indicates that the GMM robustly distinguishes the Coulomb phase from the String phase, even in the presence of string-like defects [see inset (vi) in Fig. 1(a)].

To construct the phase diagrams, for a given $Pe$ and $\phi$, we compute the probability $p_i$ that the corresponding data point belongs to class $i$.
Specifically, $p_i$ is evaluated as the fraction of samples assigned to class $i$ based on the posterior probabilities obtained from the GMM, where $i = 1, 2, 3$ correspond to the Cluster, String, and Coulomb phases, respectively.
Figures~\ref{fig:phase_RDF}(e)-(g) show heat maps of $p_i$, together with the phase boundaries obtained from the RDF analysis, represented as scattered points.
These results show that each $p_i$ varies gradually across parameter space rather than exhibiting sharp boundaries.
This behavior indicates that our method captures the crossover nature of the nonequilibrium phase transitions, where structural properties change continuously within a narrow parameter window.
Furthermore, the regions of gradual variation in $p_i$ coincide with the phase boundaries identified by the RDF analysis.
This agreement validates that our method successfully identifies the nonequilibrium phases.

To validate the phase diagram using an independent methodology, we employ an unsupervised anomaly detection \cite{Kottmann2020, Acevedo2021, Kaming2021, Kottmann2021}.
In this approach, an autoencoder is trained on a restricted region of the parameter space, and its reconstruction loss is evaluated over the entire space.
Since the autoencoder cannot accurately reproduce data with features different from those in the training region, a high reconstruction loss indicates dissimilarity from the training data.
When the training data are restricted to samples belonging to a single class, the reconstruction loss serves as an indicator of whether a test point belongs to the same class.
A key advantage of this anomaly detection approach is that it does not require a priori assumptions regarding the number of phases $K$.
Thus, this method provides an independent validation of the clustering results by quantifying dissimilarity from limited regions with $p_i \approx 1$.

Figures~\ref{fig:phase_RDF}(h)-(j) show the reconstruction loss of the autoencoder, where the training regions are indicated by pink squares.
The architecture of the autoencoder is the same as that used in our main method.
As shown in Fig.~\ref{fig:phase_RDF}(h), the region where the autoencoder trained on the $p_1$-dominant patch yields a lower loss coincides with the high-$p_1$ region in Fig.~\ref{fig:phase_RDF}(e).
The autoencoder trained on the $p_2$-dominant patch shows similar consistency, showing higher reconstruction loss at low and high $\phi$ for small $Pe$ [Fig.~\ref{fig:phase_RDF}(i)], where the system exhibits more ordered structures corresponding to the Cluster and Coulomb phases.
Finally, as shown in Fig.~\ref{fig:phase_RDF}(j), the results for the autoencoder trained on the $p_3$-dominant region are also consistent with the $p_3$ distribution shown in Fig.~\ref{fig:phase_RDF}(g).

{\em Summary.--}
We applied our autoencoder--GMM framework with Fourier preprocessing to binary colloidal systems under shear.
This preprocessing suppresses noisy features in raw configuration data that are difficult for the autoencoder to reproduce while emphasizing structural features characteristic of each phase.
As a result, the encoded data form distinct clusters in the latent space, which are robustly identified by the GMM.
Finally, the phase diagram is constructed from the fraction of samples assigned to each cluster and is found to be consistent with the RDF and anomaly-detection analyses.

Understanding the behavior of multicomponent colloidal systems under nonequilibrium conditions is important for the design of functional materials controlled by external driving forces \cite{Glotzer2007, Avishek2023, Harraq2022, Arunkumar2022, Zhang2023}.
However, the spatially heterogeneous structures induced by the competing interactions and external driving make their theoretical understanding highly challenging.
In this study, we proposed a framework that treats such complex structures through their statistical features and extracts the essential degrees of freedom characterizing the system behavior.
Our results suggest a new data-driven approach toward understanding colloidal structures under a wide variety of nonequilibrium conditions.

Our results reveal the importance of preprocessing physical data before feeding it into neural networks.
One might expect that sufficiently large neural network architectures could automatically extract appropriate representations directly from raw data, thereby eliminating the need for manual preprocessing.
However, as demonstrated in the Supplemental Material, the autoencoder–GMM framework fails when applied directly to raw configuration data, regardless of the neural network architecture \cite{SupMat}.
This failure suggests that neural networks struggle to learn highly non-local transformations, such as the Fourier transform, from scratch.
Furthermore, even if a neural network were to successfully learn an adequate representation, its black-box nature would make it difficult to extract physically meaningful information or to identify the features responsible for phase classification.
Our findings demonstrate that incorporating physically motivated representations can provide an efficient and interpretable route to analyzing complex nonequilibrium phases without relying solely on increasingly large neural-network architectures.

{\em Acknowledgments.--}
We thank Prof. Anna Dawid for valuable discussions.
Y.M. was supported by JST SPRING, Grant Number JPMJSP2139.
T.H. was supported by JSPS KAKENHI Grant Nos. JP22K13983 and JP25K01290.

\clearpage
\appendix

\onecolumngrid

\section{System}

In our system, binary colloids of diameter $\sigma$ interact through a DLVO-type potential $U_{\mathrm{DLVO}}$, which consists of screened Coulomb $U_{\text{el}}$ and van der Waals $U_{\text{vdW}}$ interactions.
The screened Coulomb interaction between two colloids at center-to-center distance $r_{ij} = |\bm{r}_i - \bm{r}_j|$ is given by
\begin{equation}
U_{\text{el}}(r_{ij}) = \pi\epsilon\sigma\phi_i\phi_j\left(1 + \frac{2\lambda}{\sigma}\right)^2\exp\left(-\frac{r_{ij} - \sigma}{\lambda}\right),
\end{equation}
where $\epsilon$ represents the solvent permittivity, $\phi_i$ is the electrostatic surface potential of the $i$th colloid ($\phi_i = \pm \phi$ with $\phi > 0$), and $\lambda$ is the Debye screening length, which is set to $\lambda = \sigma/4$.
The van der Waals interaction $U_{\text{vdW}}(r_{ij})$ can be expressed as 
\begin{equation}
U_{\text{vdW}}(r_{ij}) = -\frac{A_{ij}\sigma}{24(r_{ij} - \sigma)},
\end{equation}
where $A_{ij}$ is the Hamaker constant, which depends on the materials of the colloids. 
To incorporate excluded-volume effects, we additionally introduce the Weeks--Chandler--Andersen (WCA) potential $U_{\mathrm{WCA}}$,
\begin{equation}
    U_{\mathrm{WCA}}(r_{ij}) =
    \begin{cases}
        4k_B T
        \left[
        \left(\frac{\sigma}{r_{ij}}\right)^{12}
        -
        \left(\frac{\sigma}{r_{ij}}\right)^6
        + \frac{1}{4}
        \right]
        &
        (r_{ij} \leq 2^{1/6}\sigma), \\
        0
        &
        (r_{ij} > 2^{1/6}\sigma),
    \end{cases}
\end{equation}
where $k_B$ is the Boltzmann constant and $T$ is the temperature.

Since the overall interaction range is comparable to the particle diameter $\sigma$, the interaction becomes negligible for $r_{ij}\geq4\sigma$.
Furthermore, to avoid the divergence of the DLVO potential at contact, which can lead to unphysical aggregation, we introduce a short-distance cutoff at $r_{ij}=1.05\sigma$.
The total interaction potential is therefore given by
\begin{equation}
    U(r_{ij})=
    \begin{cases}
        U_{\mathrm{WCA}}(r_{ij})
        +
        U_{\mathrm{DLVO}}(1.05\sigma)
        &
        (r_{ij}\leq1.05\sigma), \\
        U_{\mathrm{WCA}}(r_{ij})
        +
        U_{\mathrm{DLVO}}(r_{ij})
        &
        (1.05\sigma < r_{ij}<4\sigma), \\
        0
        &
        (r_{ij}\geq4\sigma).
    \end{cases}
\end{equation}

The characteristic time scale in our simulations is defined as $t_0=\sigma^2\xi/k_BT$, where $\xi$ is the friction coefficient.
This corresponds to the characteristic time for a colloid to diffuse over a distance equal to its diameter $\sigma$.
The equation of motion is integrated using the Euler–Maruyama scheme with a time step of $\Delta t = 10^{-4} t_0$.
The interaction and simulation parameters are summarized in Table~\ref{tab:simulation_paras}.

The Hamaker constant $A_{ij}$ depends on the species of the interacting colloids and can be approximately estimated by
\begin{equation}
    A_{ij}
    =
    \left(
    \sqrt{A_i}
    -
    \sqrt{A_{\mathrm{sol}}}
    \right)
    \left(
    \sqrt{A_j}
    -
    \sqrt{A_{\mathrm{sol}}}
    \right),
\end{equation}
where $A_i$ and $A_{\mathrm{sol}}$ denote the Hamaker constants of the material composing the $i$th colloid and the solvent in vacuum, respectively \cite{Israelachvili2011}.

\begin{table}[t]
\centering
\caption{List of parameters in our simulation}
\label{tab:simulation_paras}

\begin{tabular}{cc|cc}
\hline
\begin{tabular}{l}
diameter $\sigma$ \\
(nm)
\end{tabular}
& 100
&
\begin{tabular}{l}
solvent permittivity $\epsilon$ \\
($\times10^{-10}$ F/m)
\end{tabular}
& 6.94
\\ \hline

\begin{tabular}{l}
surface potential $\phi$ \\
(mV)
\end{tabular}
& 1--20
&
\begin{tabular}{l}
temperature $T$ \\
(K)
\end{tabular}
& 298
\\ \hline

\begin{tabular}{l}
Hamaker constant of \\
positive colloids $A_p$ \\
($\times10^{-20}$~J)
\end{tabular}
& 18
&
\begin{tabular}{l}
Hamaker constant of \\
negative colloids $A_n$ \\
($\times10^{-20}$~J)
\end{tabular}
& 10
\\ \hline

\begin{tabular}{l}
Hamaker constant of \\
solvent $A_{\mathrm{sol}}$ \\
($\times10^{-20}$~J)
\end{tabular}
& 4
&
Peclet number $Pe$
& 0.1--2
\\ \hline
\end{tabular}
\end{table}

\section{annealing protocol}
To sample configurations in equilibrium states shown in the insets (i)-(iii) of Fig. 1(a), we employ the standard Monte Carlo method with an annealing protocol.
The target temperature is set to $T_0 = 297\,\mathrm{K}$, which is the same temperature used in the main simulations.
The simulation starts from a random initial configuration at $5T_0$, and the temperature is decreased by $0.04T_0$ every $2\times10^4$ Monte Carlo (MC) steps until reaching $T_0$.

At each MC step, a randomly selected colloid is displaced by $\Delta \bm{r}$, where each component of $\Delta \bm{r}$ is sampled from a normal distribution with zero mean and variance $\sigma^2$.
The trial displacement is accepted according to the Metropolis probability
\begin{equation}
  P
  =
  \min
  \left\{
  1,
  \exp
  \left(
  -\frac{\Delta E}{k_B T}
  \right)
  \right\},
\end{equation}
where $\Delta E$ denotes the energy change induced by the displacement.

\section{sampling time for configurations in the nonequilibrium phases}

\begin{figure}
\centering
\includegraphics[width=\textwidth]{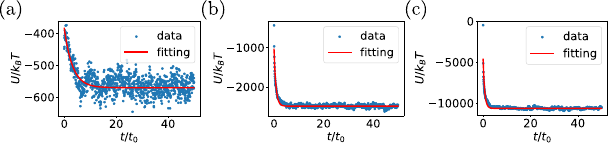}
\caption{Potential energy $U/k_B$ as a function of time $t/t_0$ for $Pe=0.1$ and $\phi=1\: \text{mV}$, $10\: \text{mV}$, and $20\: \text{mV}$, shown in panels (a), (b), and (c), respectively. The potential energy is averaged over five independent initial conditions, and solid lines represent exponential fits.}
\label{fig:potential}
\end{figure}

We determine the sampling time $t_{\text{samp}}$ for collecting configurations in nonequilibrium steady states by analyzing the relaxation time of the potential energy.
Since shear tends to accelerate the dynamics in our system, we use the parameter sets with the lowest shear rate, $Pe=0.1$, and $\phi=1,10,20\: \text{mV}$ as reference cases for determining $t_\text{samp}$.

Figures~\ref{fig:potential} show the time-dependent potential energy $U$ averaged over five different initial conditions together with exponential fitting curves.
The characteristic relaxation times obtained from the exponential fits are $3.19t_0$, $0.76t_0$, and $0.66t_0$ for $\phi=1\: \text{mV}$, $10\: \text{mV}$, and $20\: \text{mV}$, respectively.
Based on these results, we set $\text{samp} = 10 t_0$, which is sufficiently longer than the relaxation times required to reach steady states.

\section{Details of the Autoencoder Used in the Main Analysis}
In the main analysis, dimensionality reduction of the Fourier-transformed configurations is performed using an autoencoder based on a fully connected neural network (FCNN).
The network consists of input and output layers with 4608 nodes, two hidden layers with 128 nodes, and a two-dimensional latent layer.
ReLU activation functions are used in the hidden layers, whereas linear activation functions are adopted for the latent and output layers.

For each parameter set, the autoencoder is trained using 100 configurations, while an additional 50 configurations are used to evaluate the test loss.
Model parameters are optimized using Adam with a learning rate of $\eta=3\times10^{-4}$.
The loss function used for training is the mean squared reconstruction error.

\section{Loss and reconstructions}

\begin{figure}
\centering
\includegraphics[width=0.5\textwidth]{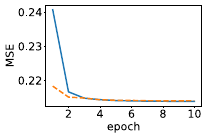}
\caption{Mean squared reconstruction error obtained from the autoencoder trained on the Fourier-transformed data. Solid and dashed lines represent the training and test errors, respectively.}
\label{fig:Fourier loss}
\end{figure}

Figure~\ref{fig:Fourier loss} presents the reconstruction error of the autoencoder trained on the Fourier-preprocessed data.
The training and test losses are evaluated using 100 and 50 independent Fourier-transformed data samples, respectively, for each parameter set.
Since no significant improvement is observed beyond 3 epochs, the network parameters obtained after 10 epochs are adopted in the main analyses.

\begin{figure}
\centering
\includegraphics[width=\textwidth]{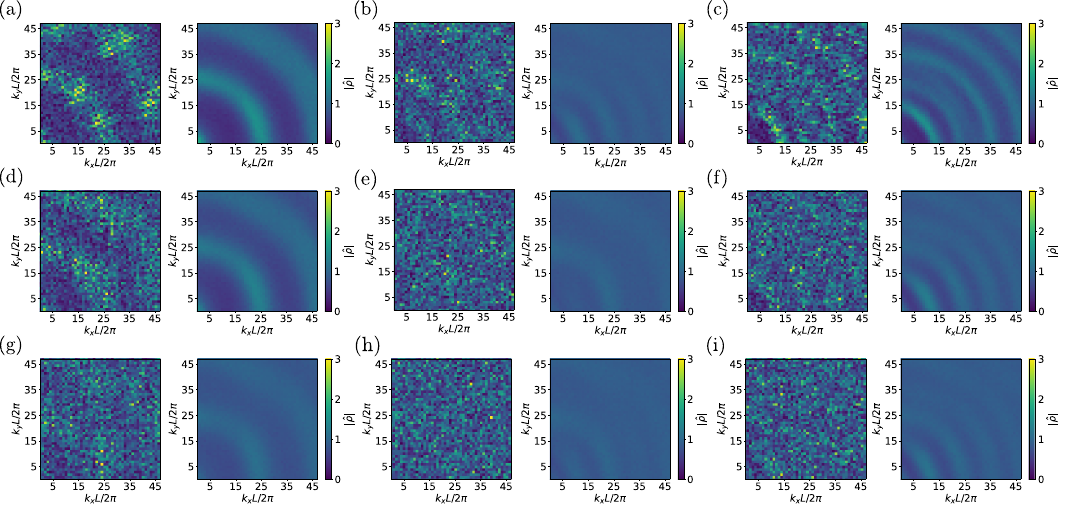}
\caption{Comparison between the original and reconstructed Fourier data. The rows correspond to $Pe=0.1$, $0.5$, and $1.5$ from top to bottom, and the columns correspond to $\phi=1\: \text{mV}$, $8\: \text{mV}$, and $20\: \text{mV}$ from left to right.
For each parameter set, the original and reconstructed data are shown on the left and right, respectively.
}
\label{fig:Fourier comparison}
\end{figure}

Figures~\ref{fig:Fourier comparison} show the comparisons between the original and reconstructed Fourier data.
Especially, Fig.~\ref{fig:Fourier comparison}(b) corresponds to the String phase.
The reconstructed Fourier data exhibit a ring with a missing upper segment at $|k| \approx 12 \times (2\pi/L)$, which is attributed to string-like structures tilted by the shear flow along the $x$ direction.
As $Pe$ increases from top to bottom, the bright ring is progressively disrupted from its upper side in both the original and reconstructed data.

\section{Failure in autoencoders' reconstruction of raw configuration}

\begin{figure}
\centering
\includegraphics[width=\textwidth]{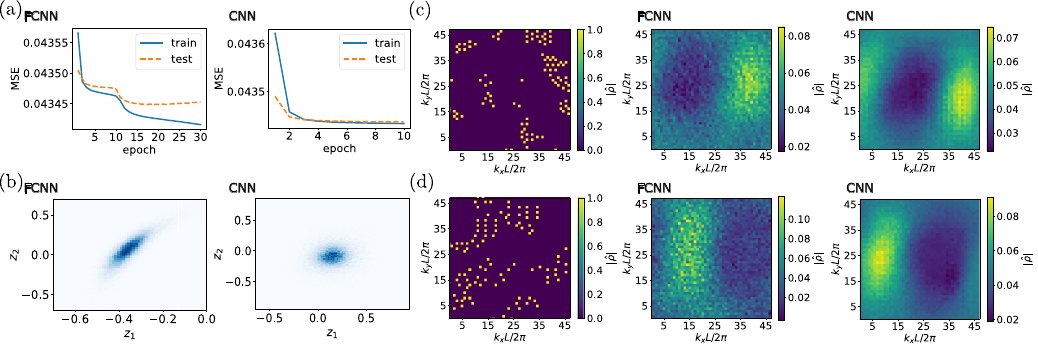}
\caption{(a) Mean squared reconstruction errors of the FCNN (left) and CNN (right) as functions of training epochs. Solid and dashed lines represent the training and test errors, respectively.
(b) Histograms of the latent space distributions obtained from the FCNN (left) and CNN (right).
(c), (d) Comparison between the original and reconstructed configurations of colloids P at $Pe=0.1$ for $\phi=1\: \text{mV}$ and $20\: \text{mV}$, respectively. In each panel, the left image shows the original configuration, and the center and right images show the corresponding reconstructions produced by the FCNN and CNN, respectively.}
\label{fig:raw data}
\end{figure}

Here, we demonstrate that raw configurations in our system are difficult for autoencoders to reconstruct accurately.
We employ both fully connected and convolutional neural networks (FCNN and CNN).
As input data, the configurations of the two colloidal species are separately discretized on grids with spatial resolution $\sigma/2$.
Each grid cell is assigned a binary value of 1 or 0, indicating the presence or absence of a colloid within that cell, respectively.
For each parameter set $(\phi,Pe)$, 100 configurations are used for training and 50 configurations are reserved for testing.
The loss function used for training is the mean squared reconstruction error.
Our FCNN consists of input and output layers with 4608 nodes, two hidden layers with 128 nodes, and a two-dimensional latent layer.
The hidden layers use ReLU activation functions, whereas the latent and output layers use linear activation functions.
Figure~\ref{fig:raw data}(a) shows the mean squared reconstruction error of our FCNN as a function of training epochs.
The test loss saturates after approximately 10 epochs, whereas the training loss continues to decrease, indicating the onset of overfitting.
Therefore, we analyze the reconstructed configurations and latent-space distribution at 10 epochs.
Figures~\ref{fig:raw data}(c) and (d) compare the original and reconstructed configurations for $\phi=1\: \text{mV}$ and $20\: \text{mV}$ at $Pe=0.1$.
The FCNN fails to reproduce the phase-specific local structures accurately, although it partially captures coarse-grained global features in high-density regions.
Consequently, the encoded data points form a unimodal distribution in latent space and fail to reflect the phase structure of the system, as shown in Fig.~\ref{fig:raw data}(b).

Our CNN encoder consists of an input layer with 4608 nodes, two convolutional layers with 20 channels, kernel sizes of $7\times7$ and $6\times6$, and stride 2, followed by ReLU activations and a two-dimensional latent layer with a linear activation function.
The decoder consists of a fully connected layer with 1280 nodes, two transposed convolutional layers with 20 channels (kernel sizes $7\times7$ and $8\times8$, stride 2), each followed by a ReLU activation, and a linear output layer with 4608 nodes.
Figure~\ref{fig:raw data}(a) presents the mean squared reconstruction error of the CNN during training.
Both the training and test losses converge within the displayed epoch range; therefore, the network trained for 10 epochs is used for further analysis.
The latent-space distribution and reconstructed configurations reveal behavior similar to that observed for the FCNN in Fig.~\ref{fig:raw data}(b), (c), and (d).
In particular, the network captures certain coarse-grained features but fails to preserve the local structural details.
This observation suggests that the strong spatial inhomogeneity of our system prevents autoencoders from extracting phase-relevant information efficiently from raw configuration data.

\bibliography{paper.bib}

\bibliography{paper.bib}

\end{document}